\documentclass[aps,apl,preprint,endfload*]{revtex4-1}

\usepackage{graphicx}
\usepackage{amssymb}
\usepackage{amsfonts}
\usepackage{amsmath}
\usepackage{amsbsy}

\renewcommand{\vec}[1]{\mbox{\boldmath$\mathrm{#1}$}}

\begin{document}

\title{Anisotropic thermoelectric effect in helimagnetic tunnel junctions}
\author{Chenglong Jia$^{1,2}$ and Jamal Berakdar$^{1}$}
\affiliation{1. Institut f\"ur Physik, Martin-Luther Universit\"at Halle-Wittenberg, 06099  Halle, Germany\\
2. Key Laboratory for Magnetism and Magnetic Materials of the Ministry of Education, Lanzhou University, Lanzhou 730000, China}

\begin{abstract}
Thermoelectric transport across normal-metal/helical-multiferroic/ferromagnetic heterojunctions is theoretically investigated. We find a anisotropic charge and spin thermopower with a $C_{2v}$ symmetry.  The angular dependence on the magnetization orientation of the ferromagnetic layer is substantiated  by a phenomenological theory based on the symmetry of the effective spin-orbit interaction induced by the topology of the spiral magnetic order in the multiferroic barrier.
\end{abstract}


\maketitle

\begin{figure}[f]
\includegraphics[width=.75\textwidth]{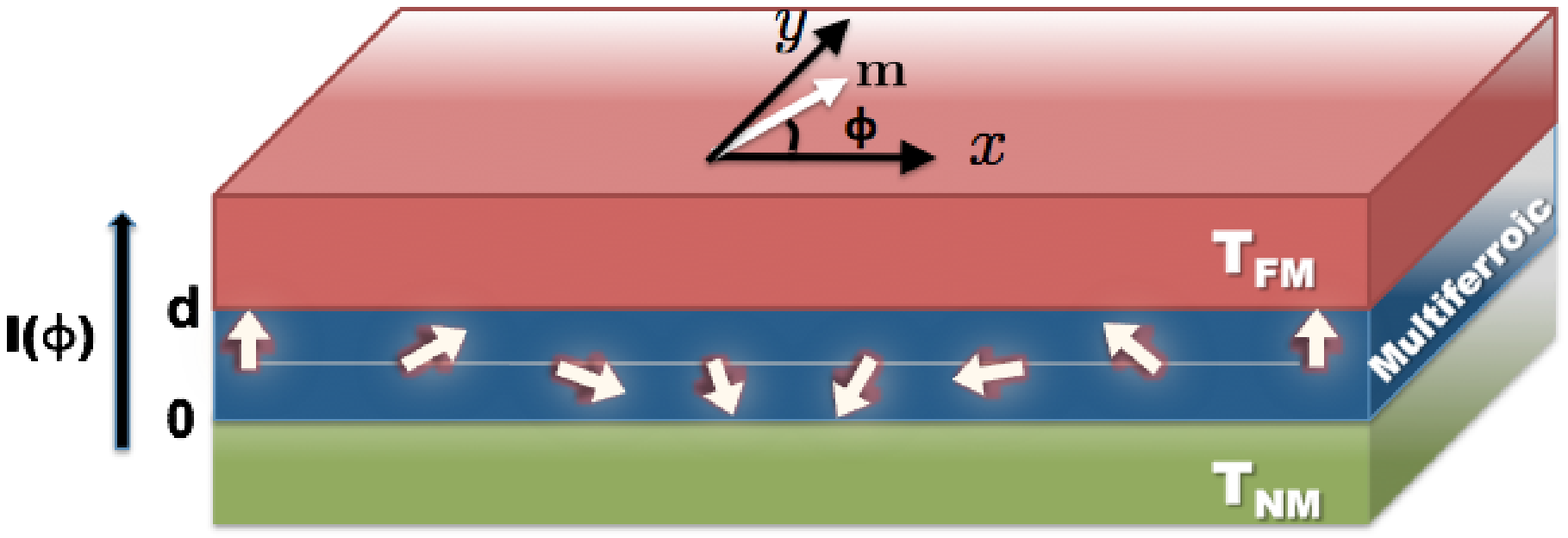}
\caption{(Color online) Schematic of a magnetic tunnel junction composed of a multiferroic oxide insulator barrier with spiral magnetic ordering (white arrows),   a normal metallic (NM) lead and ferromagnetic (FM) electrode at different temperatures, $\text{T}_{\text{NM}} \neq \text{T}_{\text{FM}}$. The vector $\vec{m}$ indicates the magnetization orientation specified by the angle $\phi$ in $xy$ (FM) plane with respected to the reference crystallographic axis ($x$-axis). The $zx$ plane refers to the spiral plane of a multiferroic oxide. }
\label{fig::MTJ}
\end{figure}

With the recent advances in fabrication, patterning  and measurement techniques of artificial nanostructures,  thermoelectricity has gained renewed interest  with
 a particular focus on potential applications \cite{Book1, Nernst-1,Nernst-2}, specially  for the development of electronic devices with low power consumption. Thermoelectric transport is well studied theoretically and experimentally for a variety of systems such as quantum dots \cite{QD-1,QD-2}, molecular junctions \cite{meJ}, and magnetic tunnel junctions \cite{TJ-1,TJ-2,Spin-Injection}.  Recently, a qualitatively original phenomena, the Spin-Seebeck effect, was discovered by Uchida \emph{et al.} \cite{SSE,SSE-M}:  In a ferromagnetic material (Ni$_{81}$Fe$_{19}$) a heat current leads to a pure spin current without a charge current. This surprising effect has by now been confirmed experimentally for insulating ferrimagnets (LaY$_{2}$Fe$_{5}$O$_{12}$ in \cite{SSE-I}), and for ferromagnetic semiconductors (GaMnAs \cite{SSE-S}) as well.  In these experiments, the spin-Seebeck coefficient was found to be quite smaller than the charge-Seebeck coefficient. However, in contrast to bulk ferromagnets, the spin thermopower in a quantum dot is theoretically predicted to be on the same order of magnitude   as the charge thermopower \cite{QD-1,QD-2}.  Along this line we  inspect  in this work the thermoelectricity in a normal-metal/helical-multiferroic/ferromagnetic heterojunctions such as Pt/TbMnO$_3$/SrRuO$_3$ (cf.  Fig.\ref{fig::MTJ}). The coexistence of coupled electric and magnetic order parameters in multiferroics \cite{RMnO3} holds the promise of futuristic opportunities for spintronics devices \cite{JB-2DEG,JB-ATMR} with the particular advantage of being electrically \cite{E-Control} and/or magnetically \cite{H-Control} controllable. The essential point is that due to the topology of the local magnetic moments in multiferroic insulator, a traversing carrier experiences an effective spin-orbit interaction \cite{JB-ATMR}. As shown by Hatimi \emph{et al.} \cite{TJ-2}, the thermoelectric effects significantly depend on the relative angle of the magnetization in neighboring magnetic layers of  the ferromagnetic/normal-metal/ferromagnetic heterojunctions. Moreover, in the presence of the spin-orbit interaction, we expect that the tunnel magnetoresistance should show a spatial anisotropy  in a ferromagnetic/insulator/normal-metal heterojunction \cite{JB-ATMR,S-spintronics,S-ATMR}. Similarly, we find that the non-collinear magnetic order in the oxide together with the induced spin-orbit coupling result in uniaxial spin and charge thermopower with a C$_{2v}$ symmetry.

A sketch of the system is shown in Fig.\ref{fig::MTJ}. A magnetic tunnel junction consisting of an ultrathin helical-multiferroic (MF) barrier (usually an oxide) is sandwiched between a normal metal (NM) lead and a ferromagnetic (FM) conductor. No voltage is applied across the junction. A steady temperature difference ($T_{NM} \neq T_{FM}$) between the FM and NM electrodes is induced. As the transmission  is spin dependent,
 the temperature gradient may lead to a spin accumulation in the electrodes, which generally results in a nonzero spin-voltage bias $\Delta V_s = (\delta \mu_{FM} - \delta \mu_{NM})/e$, where $e$ is the electron charge and $\delta \mu_i = \delta \mu_{i\uparrow} - \delta \mu_{i\downarrow}$ with $\delta \mu_{i\sigma}$ being the electrochemical potential of the spin $\sigma$ to the right or to the left of the oxide barrier. In the linear response regime (which is assumed here), the spin dependent current is written as \cite{TJ-1},
\begin{equation}
 I_{\sigma} = e L_{0\sigma} \Delta \mu_\sigma + \frac{e}{T}L_{1\sigma} \Delta T
\end{equation}
where $\Delta \mu_\sigma = e \Delta V_{\sigma}$ is the difference in the chemical potentials of the two leads in the spin channel $\sigma = \pm 1 (\text{or} \uparrow \downarrow)$, and $\Delta V_\sigma  = \Delta V_e +\sigma \Delta V_s$ with $\Delta V_e$ being the charge bias. $\Delta T$ is the applied temperature difference. The kinetic coefficients $L_{n\sigma}$ are defined as \cite{T-Landauer}
\begin{equation}
L_{n\sigma} =- \frac{1}{\hbar} \int \frac{dE}{2\pi} ~ (E- \mu)^n T_{\sigma} (E)\frac{\partial f(E)}{\partial E}
\label{Ln}
\end{equation}
where $f(E)$ is the Fermi-Dirac distribution function. $T_{\sigma} (E)$ is the spin-dependent transmission probability through the tunnel junction.
The charge and the spin current are thus given as $I_e = I_{\uparrow} + I_{\downarrow}$, $I_s = I_{\uparrow} -  I_{\downarrow}$, respectively.
We assume the spin-orbit interaction in the \emph{electrodes} to be negligible, hence the two spin channels are independent.  A nonzero spin current can be driven by the temperature difference through the system in absence of a charge current $I_e =0$ as observed in the spin-Seebeck effect \cite{SSE}. In the present study, we derive the thermoelectric and thermospin coefficients in the presence of the spin accumulation.
To find  the spin-dependent Seebeck coefficient $S_\sigma $,
we inspect the limit of the simultaneous  vanishing  of both the spin current and the charge current; or equivalently a zero charge current in each spin channel, i.e.  $I_\sigma =0$,
\begin{equation}
S_\sigma =\frac{ \Delta V_\sigma }{\Delta T} = -\frac{1}{eT} \frac{L_{1\sigma}}{L_{0\sigma}}.
\end{equation}
Analogously, the spin thermopower $S_s$ and the charge thermopower are calculated as,

\begin{eqnarray}
 S_s &=& \frac{\Delta V_s}{\Delta T} = \frac{1}{2} (S_\uparrow - S_\downarrow), \\
 S_e &=& \frac{\Delta V_e}{\Delta T} = \frac{1}{2} (S_\uparrow + S_\downarrow).
\end{eqnarray}

Neglecting the relatively much smaller potential modification due to  the depolarizing field in the multiferroic barrier, and assuming that the barrier potential has a rectangular shape  with the height $V_0$, the Hamiltonian describing the tunneling across the heterojunction reads \cite{JB-ATMR}, $H = H_m +H_{MF}$ .  $H_m$ stands for
 the itinerant  carriers in the two metal electrodes,
\begin{equation}
H_m = - \frac{\hbar^2}{2m_e} \vec{\nabla}^2 - \Theta(z-d) \Delta \vec{m}\cdot \vec{\sigma}
\end{equation}
where $\vec{\sigma}$ is the vector of Pauli matrices, $\vec{m} = [\cos \phi, \sin \phi, 0]$ is a unit vector defining the in-plane magnetization direction in the ferromagnet with respect to the [100] crystallographic direction, and $\Delta$ describes the Zeeman splitting in the FM electrode. $\Theta (z)$ is the Heaviside step function. $m_e$ is the free-electron mass and $d$ is the thickness of the barrier. In the oxide insulator, the carrier dynamics is governed by the exchange model,
\begin{equation}
H_{MF} = -\frac{\hbar^2}{2m} \vec{\nabla}^2 + J \vec{n_r} \cdot \vec{\sigma} + V_0, ~~\mbox{for}~~ 0\leq z \leq d
\end{equation}
where $m$ is the effective electron mass of the oxide ( $m/m_e \approx 3$).
$J\mathbf{n_{r}}$ is the exchange field, where $\mathbf{n_{r}}$ is given by the multiferroic oxide local magnetization at each spiral layer (labeled by the integer number $l$) along the $z$-axis \cite{helicity}, i.e., $\mathbf{n_{r}} = (-1)^{l} [ \sin \theta_r, 0, \cos \theta_r]$ with $\theta_r = \mathbf{q}_m \cdot \mathbf{r}$ and $\mathbf{ q}_m = [q,0,0]$ being the spiral spin-wave vector.  In effect the exchange coupling acts on the electron as a non-homogenous magnetic field. Performing a local unitary transformation within the barrier \cite{JB-2DEG}, we conclude
 that the influence of the barrier amounts to the spin-dependent potential
\begin{equation}
H_{SO}^{eff} = \vec{w}(\theta_r,\vec{k}) \cdot \vec{\sigma}
\label{SOI}
\end{equation}
where
\begin{equation}
\mathbf{w}(\theta_r,\mathbf{k}) = [J(z) \sin \theta_r , \tilde{q} k_x, J(z) \cos \theta_r]\end{equation} and $\tilde{q}= \frac{\hbar^2}{2m} q.$ This effective spin-orbit interaction results in a tunneling anisotropic magnetoresistance (TAMR) effect\cite{JB-ATMR}. Hence, we can expect a similar anisotropic behavior of the thermopower, as well.

\begin{figure}[f]
\includegraphics[width=.8\textwidth]{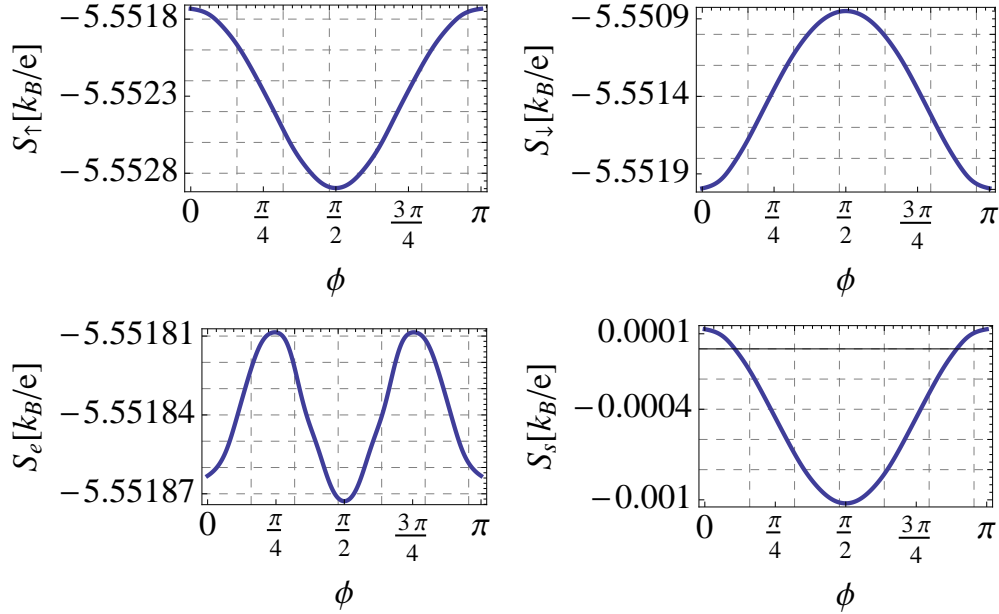}
\caption{(Color online) Spin and charge thermopower as a function of the magnetization orientation $\phi$ in FM layer at  $k_B T = 4meV$. Other parameters are chosen as $\mu =5.0 eV$, $\Delta = 2 eV$, $V_0 = 0.5 eV$, $d = 1$nm, $\bar J = 1 eV$ and $q = \frac{2\pi}{7a}$ with $a=5$\AA  being the lattice constant of the oxide.}
\label{fig::thermopower}
\end{figure}

In the present study, we assume a barrier of two to five layers (as in \cite{DDFM})
such that
 the effective spin-orbit interaction $H_{MF}^{\sigma}$ throughout the multiferroic barrier is reduced
 to the plane of the barrier, $\bar{H}_{MF}^{\sigma} = \bar{\mathbf{w}}(\theta_r,\mathbf{k}) \cdot \vec{\sigma} \delta(z)$ with $\bar{\mathbf{w}}(\theta_r,\mathbf{k}) = [\bar{J} \sin \theta_r, \bar{q} k_x, \bar{J} \cos \theta_r]$. $\bar{J}$ and $\bar{q}$ are renormalized exchange and resonant spin-orbit coupling parameters, $\bar{q} \approx qd m_e/m$ and $\bar{J} \approx \langle J(z) \rangle_d$ referring to space and momentum averages with respect to the unperturbed states at the Fermi energy. In the following, we treat $\bar{J}$ and $\bar{q}$ as adjustable parameters. The transmissivity of a spin-$\sigma$ electron through the multiferroic tunnel junctions reads
\begin{equation}
T_{\sigma}(E, \mathbf{k}_{\|},\theta_r) = \Re\left[\frac{k_{\sigma}}{\kappa} |t_{\sigma,\sigma}|^2 + \frac{k_{\bar{\sigma}}}{\kappa} |t_{\sigma,\bar{\sigma}}|^2 \right]
\end{equation}
where the transmission ($t_{\sigma,\sigma}$ and $t_{\sigma,\bar{\sigma}}$) coefficients can be analytically obtained by solving for the scattering states in the different regions \cite{JB-ATMR}. $\kappa$ and $k_\sigma$ are the transverse wave vectors in NM and FM subsystems, respectively,
\begin{eqnarray}
&&\kappa = \sqrt{ E/\frac{\hbar^2}{2m_e} -k_{\|}^2  }, \\
&& k_{\sigma} = \sqrt{ (E + \sigma \Delta)/\frac{\hbar^2}{2m_e} -k_{\|}^2  },
\end{eqnarray}
where $\mathbf{k}_{\|}$ denotes the conserved electron momentum parallel to the junction interfaces.

Introducing the transmissivity $T_{\sigma}(E, \mathbf{k}_{\|},\theta) $ into Eq.(\ref{Ln}), the kinetic coefficients are rewritten as,
\begin{equation}
L_{n\sigma} (\phi) =- \frac{1}{h} \int dE \frac{d^2 \vec{k}_{\|}}{(2\pi)^2} \frac{d\theta}{2\pi}~ (E- \mu)^n T_{\sigma} (E,\mathbf{k}_{\|},\theta)\frac{\partial f(E)}{\partial E}
\end{equation}
Based on a general symmetry considerations of the spin-orbit interaction\cite{pTAMR} and phenomenological calculations \cite{JB-ATMR} the  angular-dependence of $L_{n\sigma}(\phi)$ is found  to exhibit a two fold symmetry, $\sim \cos 2\phi$.
Consequently, the charge and spin thermopower are spatially anisotropic.

We performed numerical calculations for $k_B T = 4meV$ with $\mu =5.0 eV$, $\Delta = 2 eV$, $V_0 = 0.5 eV$, and $d = 1$nm. Fig.\ref{fig::thermopower} presents the dependence of the thermopower on the magnetization direction in the FM electrode.  As evident from the numerical results, the spin and the charge thermopower show the $C_{2v}$ symmetry, as  follows from the phenomenological model. We note,  the spin thermopower $S_s$ is about three orders of magnitude smaller than the charge thermopower $S_e$, which is different from the case of a quantum dot where $S_s$ can be as large as $S_e$ \cite{QD-1}.  $S_s$ changes sign as voltage induced in the minority spin channel is higher.  For the charge thermopower, the amplitude of the angular-dependence of $S_e$ is quite small,  which is on the same order as the tunnel anisotropic magnetoresistance in Fe/GaAs/Au tunnel junction that have been recently realized experimentally \cite{TAMR}. However, $S_s$ changes with $\phi$ from positive to negative, we have a quite large tunnel anisotropic spin thermopower, $[(S_{s}(\phi) - S_{s}(0))/(S_{s}(\phi) + S_{s}(0))]_{max} \approx 18$.

Summarizing, we studied the angular-dependence of the thermoelectric transport through the helical-multiferroic tunnel junctions, both the spin and charge thermopower are found to exhibit an anisotropic behavior due to the spiral magnetic order together with an induced spin-orbit interaction in the multiferroic spacer.  Based on the magnetoelectric coupling, the strength of the effective spin-orbit coupling is electrically/magnetically controllable and thus the spin and charge thermopower in the helimagnetic tunnel junctions. For practical applications a multilayer configuration
might be more appropriate to enhance the effect. 

This work is supported by the German Science Foundation, DFG through  SFB762 -B7- {\it functionality of oxide interfaces}.

\newpage

\end{document}